\documentclass[twocolumn,showpacs]{revtex4}

\usepackage{graphicx}
\usepackage{dcolumn}
\usepackage{amsthm}
\usepackage{amsmath}
\usepackage{amssymb}
\usepackage{mathrsfs}

\begin{document}

\title{Relative Spin Polarization of Parity-Violating Asymmetries}

\author{J.C. Yoon}
\email{jcyoon@u.washington.edu}

\date{\today}

\begin{abstract}
Parity-violating asymmetries in polarized electron scattering have
been interpreted as the asymmetries between opposite helicities of incoming fermion based on the approximation of the spin polarization operator. Here exact calculations of cross sections for parity-violating asymmetries in SLAC E158 and SLD have been performed using spin projection operators. And the parity-violating factor incorporating with spin polarization and momentum has been identified and shown that its sign depends on the spin polarization of incoming particle and the relative velocity of incoming and target particles. Therefore, I suggest a new concept of relative spin polarization to interpret the parity-violating asymmetry as contributed by the antisymmetric nature of the weak interactions depending on whether the spin direction of the incoming electron is inward or outward relative to the target electron.
\end{abstract}

\maketitle

\section{Introduction}
The parity-violating asymmetries of SLAC E158 and SLD experiments
have been interpreted as the asymmetries between opposite
helicities for the Standard
Model~\cite{Glashow,Weinberg,Salam,Prescott:1978tm,SLACE158,SLD2003}
using the chiral projection, approximated from the spin projection
operator.

Helicity $h\pm$ is defined by the momentum of a fermion and its
spin orientation; if the spin orientation is in the same direction
as its momentum, it is called right-handed helicity($h+$). And
helicity can be observed to be opposite depending on the reference
frame and is especially indefinite in the rest frame where
momentum is zero. For example, a right-handed helicity massive
fermion $u_{h+}(+p_{z}, +s_{z})$ in one reference frame can be
observed at the same time as $u_{h-}(-p_{z}, +s_{z})$ with
left-handed helicity by observers in other reference frames.
Chirality (L,R) is defined to indicate either of the two-component
objects in a massive fermion field, and no physical measurement is
available for the chirality of massive fermions and only when
fermions are massless chirality becomes helicity.

After a review of spin polarization, we investigate the validity
of the chiral projection as an approximation of the spin
polarization operator in the practical calculation of matrix
elements~\cite{PeskinSchroeder,BjorkenDrell,JCYoon}.

\section{Spin Polarization and Chiral Projection}

Spin polarization is defined by the direction of the spin relative
to a given coordinate system, whereas helicity is relative to the
momentum direction of the particle. The spin polarization vector
$s_{\mu}$ can be derived from the spin projection operator
$P(\mathbf{s})$ for an electron with $\mathbf{s}$ being the spin
direction at rest:
\begin{eqnarray}
P(\mathbf{s}) &\equiv& c u \overline{u} \nonumber \\
&=& {1 \over 4}\left(1 + {p\!\!\slash \over m}\right)\left(1 -
\gamma^{5} {{s}\!\!\!\slash}\,\right)
\end{eqnarray}
where the spin polarization 4-vector $s_{\mu}$ is
\begin{eqnarray}
s_{\mu} = \left( {\mathbf{s} \cdot \mathbf{p} \over m}, \mathbf{s}
+ {\mathbf{p}(\mathbf{s} \cdot \mathbf{p}) \over m (E + m)}
\right),
\end{eqnarray}
satisfying that
\begin{eqnarray}
s^{2} = -1 \qquad \mathrm{and} \qquad s^{\mu}p_{\mu} = 0,
\end{eqnarray}
where it is normalized to one particle per unit volume in the rest
frame by the normalization factor $c$
\cite{OkunLQ,OkunWI,Tolhoek}. For example, the spin projection
operator for an electron with the spin direction $+s_{z}$ and
momentum $+p_{z}$ is:
\begin{eqnarray}
P(+s_{z}) &\equiv& c u_{h+}(+p_{z},+s_{z})
\overline{u}_{h+}(+p_{z},+s_{z}) \nonumber \\
&=& {1 \over 2m }\left( \begin{array}{cccc} m & 0 & E-p_{z} & 0 \\
0 & 0 & 0 & 0 \\ E+p_{z} & 0 & m & 0 \\ 0 & 0 & 0 & 0
\end{array} \right)
\nonumber \\
&=& {1 \over 4}\left(1 + {p\!\!\slash \over m}\right)\left(1 -
\gamma^{5} {{s}\!\!\!\slash}\,\right)
\end{eqnarray}
and the spin polarization vector is:
\begin{eqnarray}
s_{\mu} &=& \left( {+p_{z} \over m}, 0, 0, 1 + {(+p_{z}) p_{z}
\over m (E + m) }\right) \nonumber \\
&=& \left( {+p_{z} \over m}, 0, 0, {E \over m}\right)
\end{eqnarray}
In general, the spin projection operators of a fermion with $\pm
s_{z}$ can be given by:
\begin{eqnarray}
P(+s_{z}) &\equiv& c u_{h\pm}(\pm p_{z},+s_{z})
\overline{u}_{h\pm}(\pm p_{z},+s_{z}) \nonumber \\
&=& {1 \over 2m }\left( \begin{array}{cccc} m & 0 & E \mp p_{z} & 0 \\
0 & 0 & 0 & 0 \\ E \pm p_{z} & 0 & m & 0 \\ 0 & 0 & 0 & 0
\end{array} \right)
\nonumber \\
P(-s_{z}) &\equiv& c u_{h\mp}(\pm p_{z},-s_{z})
\overline{u}_{h\mp}(\pm p_{z},+s_{z}) \nonumber \\
&=& {1 \over 2m }\left( \begin{array}{cccc} 0 & 0 & 0 & 0 \\
0 & m & 0 & E \pm p_{z} \\ 0 & 0 & 0 & 0 \\ 0 & E \mp p_{z} & 0 &
m
\end{array} \right)
\end{eqnarray}
Note that the location of the non-zero matrix elements is
determined by the spin direction from the product of the
two-component spinors $\xi \xi^{\dagger}$.

The approximation of the spin projection operator $P(\mathbf{s})$
should be performed after the full evaluation of the matrix
element, taking into account the normalization factor. For
example, the trace of the matrix element calculation
$\mathrm{tr}[{P(+s_{z})}] = 1$ vanishes for the massless
approximation ($m \rightarrow 0$) ignoring the normalization
factor $1 / 2m$:
\begin{eqnarray}
\mathrm{tr}[{P(+s_{z})}] \rightarrow {1 \over 2m }\left(
\begin{array}{cccc} 0 & 0 & E-p_{z} & 0 \\
0 & 0 & 0 & 0 \\ E+p_{z} & 0 & 0 & 0 \\ 0 & 0 & 0 & 0
\end{array} \right) \rightarrow 0.
\end{eqnarray}
The common practice of spin polarization approximation is
inaccurate compared with the accurate evaluation of the full spin
projection operator. For the relativistic limit of $m/E
\rightarrow 0$, the spin projection operators for longitudinally
polarized electrons with $\mathbf{s}$ parallel to $\mathbf{p}$ can
be reduced as:
\begin{eqnarray}
s^{\mu} &=& {1 \over m \beta}p^{\mu} - {\sqrt{1 - \beta^{2}} \over
\beta}g^{\mu 0} \nonumber \\
&\rightarrow& {p^{\mu} \over m}  \qquad \qquad \textrm{for} \;\;
\beta \rightarrow 1,
\end{eqnarray}
whereas the explicit expression of the spin polarization
${P(+s_{z})}$ is now given by:
\begin{eqnarray}
{P(+s_{z})} &\rightarrow& {1 \over 4}\left(1 + {p\!\!\slash \over
m}\right)\left(1 - \gamma^{5} {{{p}\!\!\!\slash} \over
m}\,\right)\nonumber \\
&=& {1 \over 4m }\left(
\begin{array}{cccc} m & 0 & 2(E-p_{z}) & 0 \\
0 & m & 0 & 2(E+p_{z}) \\ 0 & 0 & m & 0 \\ 0 & 0 & 0 & m
\end{array} \right)
\end{eqnarray}
and although the approximation holds true for some specific
calculations such as $\mathrm{tr}[{P(+s_{z})}] = 1$ it is
inconsistent with the accurate evaluation of the matrix element in
general. The omission of the normalization factor is also
misleading because it neglects the electron mass term in the
relativistic limit \cite{LampeReya}:
\begin{eqnarray}
{1 \over 2}\left({p\!\!\slash } + m \right)\left(1 - \gamma^{5}
{{s}\!\!\!\slash}\,\right) \rightarrow {1 \over
2}\left({p\!\!\slash } \right)\left(1 - \gamma^{5}
{{s}\!\!\!\slash}\,\right).
\end{eqnarray}
whereas the contribution of the $m$ term to the matrix element
should be the same regardless of the energy $E$ once the proper
normalization is considered, since the mass factor in $ms^{\mu}$
is normalized to be one by $m^{-1}$. The simplification of the
matrix element calculation in the relativistic limit can be given
by:
\begin{eqnarray}
\left( {{1 \pm \gamma^{5} {s \!\!\slash}_{h+}} \over 2
}\right)\left( {{p\!\!\slash + m} \over 2m}  \right) \rightarrow
\left( {{1 \pm \gamma^{5}} \over 2 }\right)\left( {{p\!\!\slash +
m} \over 2m}  \right), \label{EqExactApprox}
\end{eqnarray}
whereas the explicit expression of the spin polarization
${P(+s_{z})}$ is now given by:
\begin{eqnarray}
\left( {{1 + \gamma^{5}} \over 2 }\right)\left( {{p\!\!\slash + m}
\over 2m}  \right) = {1 \over 2m } \left(
\begin{array}{cccc} m & 0 & E-p_{z}  & 0 \\
0 & m & 0 & E+p_{z} \\ 0 & 0 & 0 & 0 \\ 0 &  0 & 0 & 0
\end{array} \right) \nonumber
\end{eqnarray}
This is inconsistent with the accurate evaluation of the general
matrix element calculation, since the spin direction represented
by a nonzero matrix element is inconsistent for massive electrons.
The matrix element for a right-handed chirality relativistic
electron to scatter to a left-handed chirality electron with
interaction $\gamma^{\mu}$ is proportional to:
\begin{eqnarray}
\lefteqn{\overline{u}_{L}(p')\gamma^{\mu}u_{R}(p)}\nonumber \\
&& {} = \overline{u}(p')\left({{1 + \gamma_{5}} \over 2}
\right)\gamma_{\mu}\left( { {1 + \gamma_{5}} \over 2}
\right)u(p)
\rightarrow 0
\end{eqnarray}
whereas from the exact representation of massive fermions in terms
of helicity, not of chirality, the corresponding matrix element is
in general nonzero:
\begin{eqnarray}
\lefteqn{\mathrm{tr}\left[\overline{u}_{h-}(p')\gamma^{\mu}u_{h+}(p)
\overline{u}_{h+}(p)\gamma^{\nu}u_{h-}(p') \right]} \nonumber \\
&&{} = \mathrm{tr}\left[\overline{u}_{h-}(p')\gamma^{\mu}u(p)
\overline{u}(p)\left({1 - \gamma^{5} {{s}\!\!\!\slash} \over 2}
\right)\gamma^{\nu}u(p')_{h-} \right]. \nonumber
\end{eqnarray}
Note that $(1 \pm \gamma^{5} {{s}\!\!\!\slash})/2$ is supposed to
act on $({p\!\!\slash + m})/2m$, not on $u$ alone, and the chiral
projection $(1 \pm \gamma^{5})$ on $u$ is inconsistent with being
a proper approximation of $(1 \pm \gamma^{5} {{s}\!\!\!\slash})/2$
for a massive fermion.

Since the chiral projection as an approximation of spin projection
operators could differ from the exact calculation of spin projection operators, the matrix element should be carefully evaluated comparing with the full representation of the spin projection operators.

\section{The Parity-Violating Asymmetry in SLAC E158 and SLD}

The parity-violating asymmetry for polarized electrons scattering
on an unpolarized target is given by:
\begin{eqnarray}
A_{PV} &\equiv& {{d\sigma_{h+}(+p_{z},+s_{z}) -
d\sigma_{h-}(+p_{z},-s_{z})} \over {d\sigma_{h+}(+p_{z},+s_{z}) +
d\sigma_{h-}(+p_{z},-s_{z})}}
\end{eqnarray}
where $d\sigma_{\lambda}$ denotes the differential cross section
for an incoming electron of helicity $\lambda$ on an unpolarized
target. For the parity-violating asymmetry in SLAC E158 and
SLD~\cite{SLACE158,SLD2003}, left- and right-handed helicity
massive fermions are considered as distinguishable particles with
the weak interaction structure of $\gamma^{\mu}(v - a\gamma^{5})$,
as suggested in the Standard Model~\cite{Swartz,Derman,Clarke},
and the asymmetry is accounted for as parity violation between
opposite helicities. For the electron scattering of
\begin{eqnarray}
e^{-}(p_{1},\lambda_{1}) + e^{-}(p_{2},\lambda_{2}) \rightarrow
e^{-}(p_{3},\lambda_{3})+ e^{-}(p_{4},\lambda_{4}),
\end{eqnarray}
where $(p_{i},\lambda_{i})$ denote the four momenta and helicities
of incoming and outgoing electrons, respectively, the interaction
Lagrangian for the electromagnetic and weak interactions with the
Z boson is given by:
\begin{eqnarray}
\mathcal{L}_{int} &\sim&
-g_{\gamma}\overline{e}\gamma_{\mu}eA^{\mu}
\nonumber \\
&+& g_{z}\bigg[c_{L}\overline{e}\gamma_{\mu}(1 - \gamma_{5})e +
c_{R}\overline{e}\gamma_{\mu}(1 + \gamma_{5})e \bigg]Z^{\mu}
\nonumber \\
&=& -g_{\gamma}\overline{e}\gamma_{\mu}eA^{\mu} +
g_{Z}\overline{e}\gamma_{\mu}(v - a\gamma_{5})eZ^{\mu},
\label{WeakLagEq}
\end{eqnarray}
where $g_{\gamma} = e$, $g_{z} = {e [2 \cos \theta_{W} \sin
\theta_{W}]^{-1}}$ and $v \equiv c_{L} + c_{R} = {1 / 2} - 2
\sin^{2} \theta_{W}$, $a \equiv c_{L} - c_{R} = {1 / 2}$. Note
that $(1 \pm \gamma^{5})$ in Eq.~(\ref{WeakLagEq}) are not chiral
projections but chiral interaction structures, and electrons $e$
are later to be classified as left- and right-handed helicity
electrons $e_{\lambda}$ using the spin polarization projection
operator $(1 \mp \gamma^{5}{{s_{1}}\!\!\!\!\!\slash}\,)$
\cite{BjorkenDrell, OkunLQ, OkunWI, Tolhoek,Derman}.

The total tree-level amplitude for $ee$ scattering via $\gamma$
and $Z$ exchange in the center-of-mass (CM) frame is then given
by:
\begin{eqnarray}
\lefteqn{\mathcal{M} = \mathcal{M}_{\gamma}^{d} +
\mathcal{M}_{\gamma}^{c}+ \mathcal{M}_{Z}^{d}+\mathcal{M}_{Z}^{c}
} \nonumber \\
&&{}= {- g^{2}_{\gamma} \over ys}
\overline{u}(p_{3})\gamma_{\mu}u(p_{1}) \cdot
\overline{u}(p_{4})\gamma^{\mu} u(p_{2}) \nonumber
\\
&&{} +{ g^{2}_{\gamma} \over (1 -
y)s}\overline{u}(p_{4})\gamma_{\mu}u(p_{1})\cdot
\overline{u}(p_{3})\gamma^{\mu}u(p_{2}) \nonumber \\
&&{}- {  g^{2}_{z} \over m^{2}_{z}}
\overline{u}(p_{3})\gamma_{\mu}(v-a\gamma_{5}) u(p_{1}) \cdot
\overline{u}(p_{4})\gamma^{\mu}(v-a\gamma_{5}) u(p_{2}) \nonumber
\\
&&{} + {  g^{2}_{z} \over m^{2}_{z}}
\overline{u}(p_{4})\gamma_{\mu}(v-a\gamma_{5}) u(p_{1})  \cdot
\overline{u}(p_{3})\gamma^{\mu}(v-a\gamma_{5}) u(p_{2}) \nonumber
\end{eqnarray}
where $y \equiv - {(p_{1}-p_{3})^{2}/ s} = \sin^{2}
(\theta_{cm}/2)$ with the CM scattering angle $\theta_{cm}$ and
the CM energy $\sqrt{s}=[(p_{1}+p_{2})^{2}]^{1/2}$.

For the explicit spin polarization calculation with the operator
$(1 \mp \gamma^{5}{{s_{1}}\!\!\!\!\!\slash}\,)$ for inward and
outward relative spin polarization corresponding to $(1 \pm
\gamma^{5})$, the nonzero parity-violation asymmetry arises from
such terms as:
$[\mathcal{M}_{\gamma}^{d}\mathcal{M}_{z}^{d*}]'_{\lambda_{1}}
+[\mathcal{M}_{\gamma}^{d}\mathcal{M}_{z}^{c*}]'_{\lambda_{1}}
+[\mathcal{M}_{\gamma}^{c}\mathcal{M}_{z}^{d*}]'_{\lambda_{1}}
+[\mathcal{M}_{\gamma}^{c}\mathcal{M}_{z}^{c*}]'_{\lambda_{1}}$,
\begin{eqnarray}
\lefteqn{[\mathcal{M}_{\gamma}^{d}\mathcal{M}_{z}^{d*}]'_{\lambda_{1}}
}\hspace{3.3in} \nonumber \\
\lefteqn{={ g_{\gamma}^{2} g_{z}^{2} \over ysm_{z}^{2}}
\mathrm{tr}[\overline{u}(p_{3})\gamma^{\mu}(1 \mp
\gamma^{5}{{s_{1}}\!\!\!\!\!\slash}\,)u(p_{1})
\overline{u}(p_{1})\gamma^{\nu}(v - a
\gamma^{5})u(p_{3})]}\hspace{3.1in}
\nonumber \\
\lefteqn{ \mathrm{tr}[\overline{u}(p_{4})\gamma_{\mu}u(p_{2})
\overline{u}(p_{2})
\gamma_{\nu}(v - a\gamma^{5})u(p_{4})]}\hspace{2.7in} \nonumber \\
\lefteqn{= [\mathcal{M}_{\gamma}^{d}\mathcal{M}_{z}^{d*}] \mp 32
{g_{\gamma}^{4} \over y}\beta  (-E_{1}p_{2z}+E_{2}p_{1z}) v a
}\hspace{3.1in}
\end{eqnarray}
\begin{eqnarray}
\lefteqn{[\mathcal{M}_{\gamma}^{d}\mathcal{M}_{z}^{c*}]'_{\lambda_{1}}
}\hspace{3.3in} \nonumber \\
\lefteqn{={ g_{\gamma}^{2} g_{z}^{2} \over ysm_{z}^{2}}
\mathrm{tr}[\overline{u}(p_{3})\gamma^{\mu}(1 \mp
\gamma^{5}{{s_{1}}\!\!\!\!\!\slash}\,)u(p_{1})
\overline{u}(p_{1})\gamma^{\nu}(v - a
\gamma^{5})u(p_{4})}\hspace{3.1in}
\nonumber \\
\lefteqn{ \overline{u}(p_{4})\gamma_{\mu}u(p_{2})
\overline{u}(p_{2})
\gamma_{\nu}(v - a\gamma^{5})u(p_{3})]}\hspace{2.45in} \nonumber \\
\lefteqn{= [\mathcal{M}_{\gamma}^{d}\mathcal{M}_{z}^{c*}] \mp 32
{g_{\gamma}^{4} \over y}\beta (-E_{1}p_{2z}+E_{2}p_{1z}) v a
}\hspace{3.1in}
\end{eqnarray}
where the incoming and target electrons are given by $p_{1} =
(E_{1}, p_{1z})$ and $p_{2} = (E_{2}, p_{2z})$. The
parity-violating asymmetry $A_{PV}$ is given by:
\begin{eqnarray}
A_{PV} &=& - 16 \beta (-E_{1}p_{2z}+E_{2}p_{1z}) v a {y(1-y) \over
1 + y^{4} + (1-y)^{4}} \nonumber
\end{eqnarray}
where $(-E_{1}p_{2z}+E_{2}p_{1z}) = E_{1} E_{2} (\vec{v}_{1} -
\vec{v}_{2}) = \sqrt{(p_{1} \cdot p_{2}) - m_{1}^{2}m_{2}^{2}}$ is
Lorentz invariant holding for any arbitrary frame when
$\vec{v}_{1}$ is parallel to $\vec{v}_{2}$ and the massless
approximation of $A_{PV}$ in the CM frame $p_{1} \simeq (E, E)$
and $p_{2} \simeq (E, -E)$ is consistent with the approximated
parity-violating asymmetry of the SLAC E158 denoted as
$A^{approx}_{PV}$ \cite{Derman}.

In SLD, the polarized differential cross section of the $e^{-}_{h\pm}e^{+} \rightarrow Z^{0} \rightarrow f \overline{f}$ process with longitudinally polarized electrons and unpolarized positrons is
\begin{eqnarray}
\lefteqn{
\bigg[{d \sigma^{Z}_{f} \over d \Omega}\bigg]_{h\pm}
= {N_{c} \alpha^{2} \over 8 s \sin^{4} \theta_{W} \cos^{4} \theta_{W} [(s-m^{2}_{Z})^{2} + s^{2}\Gamma^{2}_{Z}/m^{2}_{Z}]} }\hspace{3.3in}
\nonumber \\
\lefteqn{\times [2(v_{e}^2+a_{e}^2)(v_{f}^2+a_{f}^2)(E_{1}E_{2}E_{4}E_{3}
+p_{1z}p_{2z}p_{4z}p_{3z})}\hspace{3.in}
\nonumber \\
\lefteqn{\mp2v_{e}a_{e}(v_{f}^2+a_{f}^2)(2E_{1}p_{2z}p_{4z}p_{3z}
+2p_{1z}E_{2}E_{4}E_{3}}\hspace{2.9in} \nonumber \\
-(E_{1}E_{2}+p_{1z}p_{2z})(p_{4z}E_{3}+E_{4}p_{3z}))
\nonumber \\
\lefteqn{+4v_{e}a_{e}v_{f}a_{f}(p_{1z}E_{2}-E_{1}p_{2z})
(E_{4}p_{3z}-p_{4z}E_{3})}\hspace{2.9in} \nonumber \\
\lefteqn{\mp2(v_{e}^2+a_{e}^2)v_{f}a_{f}(E_{1}E_{2}-p_{1z}p_{2z})
(E_{4}p_{3z}-p_{4z}E_{3})}\hspace{2.9in}
\nonumber \\
\lefteqn{-(v_{e}^2+a_{e}^2)(v_{f}^2+a_{f}^2)(E_{1}p_{2z}
+p_{1z}E_{2}) }\hspace{2.9in}
\nonumber \\ \lefteqn{\times (E_{4}p_{3z}+p_{4z}E_{3})
]}\hspace{1.5in}
\end{eqnarray}
Note that its massless
approximation in the CM frame $p_{1} \simeq (E, E)$
and $p_{2} \simeq (E, -E)$ is consistent with the approximated
parity-violating calculation of the SLD \cite{SLD2003} and the total cross sections are Lorentz invariant.

\begin{figure}
\includegraphics[width=0.30\textwidth,height =
0.13\textheight,viewport=100 0 450 220]{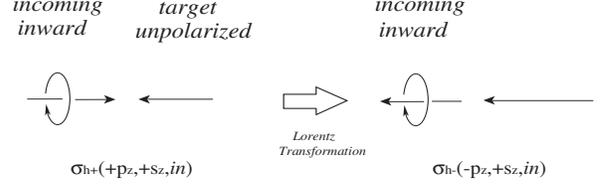} \caption{
The spin-polarized incoming electron scattering under Lorentz
transformations} \label{fig:LT2Particle}
\end{figure}

\section{Relative Spin Polarization for two-particle systems}

Let us investigate the exact cross section calculation under parity regarding to spin polarization and momentum of incoming and target particles. In the exact calculations, the factor that determines the signs of parity-violating terms depending on spin polarization and momentum is given by $\pm (E_{1}p_{2z} - E_{2} p_{1z}) = \mp E{1}E_{2}(\vec{v}_{1} - \vec{v}_{2})$. The signs of parity-violating terms become reversed when either the spin polarization ($\pm s_{z}$) of the incoming particle or the relative velocity between two particles $\vec{v}_{r} = (\vec{v}_{1} - \vec{v}_{2})$ is reversed. Note that the signs of parity-violating terms can remain constant even when the helicity of incoming electron becomes opposite as its momentum is reversed ($\vec{v}_{1} \rightarrow -\vec{v}_{1}$) under the Lorentz transformations such as in Fig.~\ref{fig:LT2Particle}, since the factor $\pm (E_{1}p_{2z} - E_{2} p_{1z}$) is Lorentz invariant along the $z$ direction.

For a two-particle system, the spin polarization can be
characterized by its direction relative to the velocity difference
between the incoming and target particles $\vec{v}_r$, indicating whether the spin direction of the incoming particle points outward from or toward the target particle. Under Lorentz transformations, the differential cross sections for a right-handed helicity incoming electron can be observed as a left-handed helicity electron, but the relative spin polarization remains inward as in Fig.~\ref{fig:LT2Particle}. Thus, the exact evaluation of parity-violating asymmetry $A_{PV}$ remains asymmetric in general:
\begin{eqnarray}
A_{PV} &\equiv& {{d\sigma_{h+}(+p_{z},+s_{z},in) -
d\sigma_{h-}(+p_{z},-s_{z},out)} \over
{d\sigma_{h+}(+p_{z},+s_{z},in) +
d\sigma_{h-}(+p_{z},-s_{z},out)}} \nonumber \\
&=& {{d\sigma_{h-}(-p_{z},+s_{z},in) -
d\sigma_{h-}(+p_{z},-s_{z},out)} \over
{d\sigma_{h-}(-p_{z},+s_{z},in) +
d\sigma_{h-}(+p_{z},-s_{z},out)}} \nonumber
\end{eqnarray}
whereas the approximated calculation of parity-violating
asymmetry, interpreted as the asymmetry between opposite
helicities, vanishes $A^{approx}_{PV}=0$.

Therefore, the parity-violating asymmetry measured in SLAC E158 and SLD should be interpreted in terms of relative spin polarization not of helicity, since the asymmetry between opposite helicities vanishes under Lorentz transformations whereas the exact calculation does not.

\section{Conclusion}

Since the approximate spin polarization $(1 \pm \gamma^{5})$ may significantly differ from the exact spin polarization in evaluating the matrix element, here exact calculations of cross sections for parity-violating asymmetries in SLAC E158 and SLD have been performed using the full expression of spin projection $(1 \mp \gamma^{5}{{s}\!\!\!\slash}\,)$. And the parity-violating factor incorporating with spin polarization and momentum $\pm (E_{1}p_{2z} - E_{2} p_{1z}) = \mp E{1}E_{2}(\vec{v}_{1} - \vec{v}_{2})$ has been identified and shown that its sign depends on the spin polarization of incoming particle and the relative velocity of incoming and target particles, not on the helicity of incoming particle. Therefore, I suggest a new concept of relative spin polarization to interpret the parity-violating asymmetry as contributed by the antisymmetric nature of the weak interactions depending on whether the spin direction of the incoming electron is inward or outward relative to the target electron.


\begin{thebibliography}
{}

\bibitem{Glashow} S.\ Glashow, Nucl.\ Phys.\ {\bf 22}, 579 (1961).

\bibitem{Weinberg} S.\ Weinberg, Phys.\ Rev.\ Lett.\ {bf 19}
(1967), 1264.

\bibitem{Salam} A.\ Salam, in {\it Elementary Particle Theory}
({\it Nobel Symposium No. 8}, ed.\ N.\ Svartholm, Stockholm, 1968)
367.
\bibitem{Prescott:1978tm}
  C.~Y.~Prescott {\it et al.},
  Phys.\ Lett.\  B {\bf 77}, 347 (1978).

\bibitem{SLACE158} P.L.\ Anthony {\it et al.}, Phys.\ Rev.\ Lett.\ {\bf 95} (2005) 081601.

\bibitem{SLD2003} Kenji Abe {\it et al.}, Phys.\ Rev.\ Lett.\ {\bf
90} (2003) 141804.

\bibitem{PeskinSchroeder} Michael E. Peskin and Daniel V.
Schoreder, {\it An introduction to Quantum Field Theory},
Addison-Wesley Publishing Co., (1995).

\bibitem{BjorkenDrell} J.D. Bjorken and S.D. Drell, {\it Relativistic Quantum
Mechanics},McGraw-Hill, New York, (1964), 141.


\bibitem{JCYoon} {For elaborate discussions with other researchers, see the website}
http://www.jcyoon.com/phpBB


\bibitem{JacobWick} M. Jacob and G.C. Wick, Ann. of Phys. {\bf
281}, (2000) 774.


\bibitem{OCPV} J.C. \ Yoon, hep-ph/0211005

\bibitem{LVSM} J.C. \ Yoon, hep-ph/0502142

\bibitem{OkunLQ} L.B. Okun, {\it Leptons and Quarks}, North-Holand Publishing Company,
(1982), 328.

\bibitem{OkunWI} L.B. Okun, {\it Weak interaction of elementary particles},
Pergamon Press, (1965), 59.

\bibitem{Tolhoek} H.A. Tolhoek, Rev. Mod. Phys., {\bf 28}, (1956) 277.

\bibitem{LampeReya} B.\ Lampe and E. \ Reya, Phys. Rept. {\bf 332} (2000) 1.

\bibitem{Streater} R.F.\ Streater and A.S.\ Wightman, {\it PCT, Spin
and Statistics, and All That}, Benjamin Cummings, Reading, 1964.


\bibitem{Clarke} P.\ Clarke, RAL-90-055 (1990).

\bibitem{Swartz} M.\ Swartz, SLAC Beam Line {\bf 25N1}, (1995),
19.

\bibitem{Derman} E.\ Derman and W.J.\ Marciano, Ann. Phys.
(Berlin) {\bf 121} (1979) 147.


\bibitem{SLACE158Mastromarino} P.A. Mastromarino {\it et al.},
IEEE Trans. Nucl. Sci. {\bf 49} (2002) 1097.

\bibitem{YoonNormal} J.C.\ Yoon, Mod. Phys. Lett. {\bf A21} (2006) 985.

\bibitem{Czarnecki} A.\ Czarnecki and W.J.\ Marciano, Phys. Rev. {\bf D53}
(1996) 1066.

\bibitem{Humensky} T.B.\ Humensky {\it et al.}, Nucl. Instrum. Meth.
{\bf A521} (2004) 261.

\end{thebibliography}
\end{document}